\documentstyle[natbib20]{mn}

\input psfig.sty

\def\Teff{\hbox{$\,T_{\rm eff}$} }
\def\kms{\hbox{$\,$km$\,$s$^{-1}$}}

\catcode`\@=11
\def\slantfrac#1#2{\hbox{$\,^#1\!/_#2$}}
 \def\ion#1#2{\hbox{#1$\,${\sc\@roman{#2}}\relax}}
\def\case#1#2{\hbox{$\frac{#1}{#2}$}}
\catcode`\@=12

\begin{document}
\pssilent

\title[Analysis of Keck High Resolution Spectra of VB10]
{Analysis of Keck High Resolution Spectra of VB10\thanks{Based on 
observations obtained at the W.M. Keck Observatory, which is operated
jointly by the University of California and the California Institute of
Technology}}

\author[A. Schweitzer et al.]{Andreas Schweitzer,$^{1,4}$ Peter H. Hauschildt,$^{1,5}$
France Allard$^2$ and G. Basri$^3$\\
$^1$ Department of Physics and Astronomy, Arizona State University, Tempe, AZ 85287-1504\\
\quad E-mail~: {\tt andy@sara.la.asu.edu \& yeti@sara.la.asu.edu} \\
$^2$ Dept.\ of Physics, Wichita State University, Wichita, KS 67260-0032\\
\quad E-Mail~: {\tt allard@eureka.physics.twsu.edu} \\
$^3$ Department of Astronomy, University of California, Berkeley, CA 94705\\
\quad E-mail~: {\tt basri@astro.berkeley.edu} \\
$^4$ present address : Landessternwarte, K{\"o}nigstuhl, D--69117 Heidelberg, Germany\\
$^5$ present address : Department of Physics and Astronomy, University of Georgia, Athens, GA 30602-2451
}

\maketitle

\begin{abstract}

We use a preliminary version of our ``NextGen'' grid of cool star model atmospheres to
compute synthetic line profiles which fit high resolution Keck spectra of the
cool M~dwarf VB10 satisfactorily well.  We show that the parameters derived from
the Keck data are consistent with the parameters derived from lower resolution
spectra with larger wavelength coverage.  We discuss the treatment of van der
Waals broadening in cool, molecular (mostly ${\rm H_2}$) dominated stellar
atmospheres.  The line profiles are dominated by van der Waals pressure
broadening and are a sensitive indicator for the gravity and metallicity.
Therefore, the high-resolution Keck spectra are useful for determining the
parameters of M dwarfs.  There is some ambiguity between the metallicity and
gravity.  For VB10, we find from the high-resolution spectra that $ 5.0 <
\log(g) < 5.5$ and $0 < \left[\case{{\rm M}}{{\rm H}}\right] < +0.5$ for an
adopted fixed effective temperature of 2700~K \cite[]{schweitzer95}, which is
consistent with recent interior calculations \cite[e.g.][]{baraffe95}.

\end{abstract}

\begin{keywords}
line: profiles -- stars: individual: VB10 -- stars: low mass, brown dwarfs --
stars: fundamental parameters -- stars: atmospheres -- atomic processes
\end{keywords}

\section{Introduction}
 
M~dwarfs are among the faintest and coolest stellar objects.  Their spectra are
dominated by molecular band absorption.  In the optical region the major opacity
sources are the TiO bands, which produce a pseudo continuum with the atomic
lines superposed.  These atomic lines show very broad wings due to van der Waals
broadening, whereas the molecular lines show only relatively small damping
wings.  We have recently obtained a high resolution optical spectrum of the dM8e
star VB10 (Gl 752B) with the HIRES echelle on the Keck telescope. The collection
and reduction of the data are as in \cite{basri95}. In this
paper, we use this spectrum to assert our treatment of the van der Waals line
broadening in M~dwarf atmospheres and to estimate the gravity and
the chemical composition of VB10.
VB10 is 
known to be a chromospherically active flare star.  However, in this
paper we will neglect the chromosphere and do not attempt to model
chromospheric spectral features, such as core emission observed in
resonance lines.

Our calculations are based on a preliminary version of our latest grid of model
atmospheres (the ``NextGen'' or ``version 5'' grid) for cool dwarf stars and
version $6.2$ of the generalized stellar atmosphere code {\sc phoenix}.  An
important improvement over the previous generation of this grid
\cite[][hereafter, AH95]{MDpap} are larger and more reliable molecular line
lists which allow us to compute detailed molecular line
profiles.  A complete description of all changes and improvements over the AH95
version of the models will be given in a subsequent paper \cite[][hereafter,
AHS96]{NGpap}.  Note that we use 
metallicities scaled from the solar values of
\cite{solab89} and we use the notation 
\[ 
\left [\frac{{\rm M}}{{\rm H}} \right ] := \log
\left[ \frac{\slantfrac{{\rm M}}{{\rm H}}}{\slantfrac{{\rm M_{\sun}}} {{\rm
H_{\sun}}}} \right]
\]
to specify abundances.

A previous analysis based on low resolution spectra by \cite{brett95}
yielded only rough values for \Teff=2400-2600 K, $\log(g)\approx 5.0$ and
$\left [\frac{{\rm M}}{{\rm H}} \right ] \approx 0.0$.
Interior calculations \cite[e.g.][]{burrow93} require for solar
M~dwarfs in the effective temperture range of $\approx 2700$ K gravities
of $\log(g) \approx 5.3$.

In the next section we will describe our theoretical approach to the problem.
We will discuss the basic approximation we use as well as their numerical
realization.  In section 3 we present the results of our modeling.  We discuss
the sensitivities of the profiles and try to establish error limits.  In
section 4 we compare the observed high resolution spectra of VB10 to our model 
spectra.

\section{The treatment of line broadening}

\subsection{The line profiles}

\begin{table*}

\caption{\label{perttab} The perturbers included in the calculations
and their polarizabilities. The percentages are taken
from  models with $\log(g)=5.0$, 
$\left[\case{{\rm M}}{{\rm H}}\right]=0.0$ and the respective ${T_{\rm eff}}$.
We chose the optical depth $\tau_{\rm std}=1\,({\rm d}\tau_{\rm std}
=\kappa{\rm d}s$
where $\kappa$ is the absorption coefficient at a standard
wavelength of {1.2$\mu{\rm m}$} (AH95) and ${\rm d}s$ the
geometric depth) as an example 
for  an average
line forming depth. 
The percentages are rounded on the last figure.
}

 \begin{center}

\begin{tabular}{llr@{.}lr@{.}lr@{.}l}
 				& 					&
				\multicolumn{6}{c}{Fraction of  ${\rm P_{gas}}$ in per cent at $\tau_{\rm std} =1$ for}\\ 
\raisebox{1.5ex}[0cm][0cm]{Perturber}	& \raisebox{1.5ex}[0cm][0cm]{$\alpha_p$ in $10^{-24} {\rm cm^3}$}	& 
				\multicolumn{2}{c}{\Teff=2000}  
						& \multicolumn{2}{c}{\Teff=2700}
								& \multicolumn{2}{c}{\Teff=3500} \\ 
\hline  
H               & 0.666793                      & 1&2	       & 14&9         & 66&6   \\ 
He              & 0.204956                      & 16&2         & 15&1         & 10&7   \\ 
Ne              & $0.3956\pm 0.1\%$             & 0&02         & 0&02         & 0&01   \\ 
Fe              & $8.4\pm 25\%$                 & 0&01         & 0&01         & \ $\approx$ 0&005\\ \hline 
${\rm H_2}$     & $0.806\pm 0.5\%$              & \ 82&4       & \ 69&8       & 22&4   \\ 
CO              & $1.95\pm 0.5\%$               & 0&06         & 0&06         & 0&04   \\ 
${\rm H_2O}$    & $1.45\pm 0.5\%$               & 0&07         & 0&06         & 0&01   \\ 
${\rm N_2}$     & $1.7403\pm 0.5\%$             & 0&01         & 0&01         & $\approx$ 0&004\\ 
\end{tabular}

 \end{center}

\end{table*}

We include all common line broadening mechanisms, i.e., thermal 
broadening, micro-turbulence, natural broadening and pressure 
broadening, both in the model calculations and synthetic spectra. The 
resulting Voigt profiles $V(u,\alpha)$, where 
$u=\Delta\lambda/\Delta\lambda_D$ and $\alpha=(\lambda^2\gamma/4\pi 
c)/\Delta\lambda_D$ ($\lambda_D$ is the Doppler width, $\gamma$ the 
Lorentz width, $\Delta\lambda$ the distance from the line center and 
$\lambda$ the wavelength), are calculated as a convolution of a Gauss 
and a Lorentz profile using standard approximations 
\cite[e.g.][]{lb}. 
The gas temperatures in M dwarf atmospheres are not high enough
to sustain a significant amount of ionization in the atmosphere.
The electron and proton densities are, therefore, much smaller than
the densities of the most important neutral and molecular species. 
Consequently, the contribution of Stark broadening to the total damping 
constant is very small, even in stars with very low metallicities.
We include a microturbulence of $\xi=2\,{\rm\, km\, sec}^{-1}$
in addition to the thermal speed of the atoms and molecules. However, 
the total thermal plus microturbulent line widths are for most lines 
much smaller than the line width due to van der Waals (vdW) broadening. 
In the following paragraphs we will describe in some detail 
the way we estimate the vdW damping constants and how their choice 
influence the computed profiles of both atomic and molecular lines.
The interaction between two different, unpolarized, neutral particles is
described by the van der Waals interaction
\begin{equation}
\Delta \nu = \frac{C_6}{r^6}
\end{equation}
where $\Delta \nu$ is the resulting frequency shift due to 
one interaction, $r$ the
relative distance between the interacting particles and
$C_6$ the interaction constant for the vdW interaction.
Within the impact or static approximation vdW damping 
leads to a Lorentz profile with a full width half maximum damping
constant $\gamma_{\rm vdW}$ (cgs-units)
\begin{equation}
\label{gammavdw}
\gamma_{\rm vdW}=17\cdot C_6^{\slantfrac{2}{5}}v^{\slantfrac{3}{5}}N_p.
\end{equation}
Here $v$ is the relative velocity between perturber and
absorber and  $N_p$ the number density
of the perturber.

We base the calculation of $C_6$ on Uns\"old's hydrogenic 
approximation \cite[]{unsold55}. However, we explicitly account for 
the different polarizabilities of each  perturber in the 
equation for $C_6$ which leads to 
\begin{eqnarray}
\label{classc6}
C_6^0&=&\frac{\alpha_p}{\alpha_{\rm H}}1.01\times 10^{-32}(Z+1)^2 
 \nonumber \\ 
& & \! \! \! \! \! \!  \times \left[ \frac{E^2_{\rm H}}{(E-E_l)^2}-\frac{E^2_{\rm H}}{(E-E_u)^2}\right]
[{\rm cm^6s^{-1}}],
\end{eqnarray}
where  $\alpha_p$ is the polarizability of the perturber,  $Z$ the 
charge of the absorber, $E$ the
ionization potential, $E_l$ the lower and $E_u$ the upper level 
excitation energy of the absorber,
$\alpha_{\rm H}$ the polarizabilty of Hydrogen, 
and $E_{\rm H}=13.6{\rm eV}$. We include the 
most abundant perturbers in M~dwarfs and their polarizabilities as
given in \cite{CRC} and listed in Tab.~\ref{perttab}
to calculate the total Van der Waals damping constant:
\begin{equation}
\label{gammatot}
\gamma^{{\rm tot}}_{\rm vdW}=\sum_{p}\gamma_{\rm vdW}^{p},
\end{equation}
where we sum over all perturbers considered.
Due to the lack of availability of any vdW broadening mechanismns for
molecules in this temperature range, we also use this approximation
for the molecular lines.

Earlier investigations \cite[]{weidemann55} showed that
the values as calculated by Eq.~(\ref{classc6}) are
in good agreement with observed line widths for alkali metals but not for
other elements, such as  
iron \cite[]{kusch58}. This has lead to
the introduction of correction factors to the `classical'
formula. Therefore, we include a correction factor
for ${\rm C_6}$ of 
\begin{equation}
\label{c6corr}
C_6=C_6^{\rm corr}\times C_6^0=10^{1.8}\times C_6^0
\end{equation}
as described in \cite{wehrselieb80} 
for non-alkali like species.
For alkali-metals and ions with alkali-like electron structure we
omit any correction factor.
This will be discussed and investigated in greater detail in sections \ref{obsfit} 
and \ref{results}. However, as these results will show, 
the use of this correction factor remains unclear for the non-alkali like species.

\subsection{The numerical treatment of vdW broadening}
\label{numeric}

In order to calculate the emerging spectrum we have to 
calculate a Voigt profile at every wavelength point for
every line at every depth point. To limit the computation
time we make the following simplification.

Before calculating the absorption coefficient,
a line selection procedure decides which lines are calculated
with Voigt profiles.
We follow the line selection procedure described by AH95.
Only lines that  are stronger than a certain threshold relative to the continuum 
are treated with Voigt profiles. This is decided by comparing the absorption
coefficient at  the line center $\kappa_l$ with the corresponding continuous 
absorption coefficient $\kappa_c$. Only if the ratio $\Gamma = \kappa_l 
/ \kappa_c$ for at least one of three representative depth points is larger
than a preset threshold value, a line is treated with a Voigt profile.
Otherwise the line is considered so weak that to a good approximation it
can be treated as a Gauss profile in order to save computation time.
The number of considered atomic lines is much smaller than the number of 
molecular lines. This allows us to set $\Gamma$ for atomic lines very low 
in order to  treat as many lines as possible with Voigt profiles.
The number of molecular lines is about two orders of magnitude larger but
since the atmosphere structure does not change significantly 
(as test calculations showed) with the inclusion of more Voigt profiles for
molecular lines we set the threshold for molecular lines for iterations to
$\Gamma = 100$.  For high resolution spectra we use lower values (usually
$\Gamma \approx 1$) although we found that the spectra do not change
significantly by increasing the number of lines with Voigt profiles.

The spectra presented throughout this paper 
used for molecular lines $\Gamma=5$ when we examined atomic lines
and $\Gamma=0.1$ when we examined molecular lines. 
All the spectra are furthermore based on self-consistent model strucutres with
thier respective parameters.

\section{Results}

\subsection{Synthetic line profiles}
\label{synprf}

\begin{figure} 
\caption{\label{diffvdw}The calculated \ion{Na}{1}~$\lambda\lambda8183,8195$
doublet in vacuum with different 
broadening mechanisms ($\log(g)=5.0$,  ${T_{\rm eff}=2700K}$, 
$\left[\case{{\rm M}}{{\rm H}}\right]=0.0$).
From top to bottom~: Gauss profile, only H and He as perturbers,
H, He and $\rm H_2$ as perturbers,
perturbers of Tab. 1 without the correction factor 
(this plot is lying exactly on the previous one), 
perturbers of Tab. 1
and  the correction factor of $10^{1.8}$ artificially applied.}
\end{figure}

\begin{figure*} 
\hbox{
}
\caption{\label{loggvdw}The calculated  \ion{Na}{1}~$\lambda\lambda8183,8195$
and \ion{K}{1}~$\lambda\lambda7665,7699$
doublets in vacuum
for $\log(g)=4.5$ (upper line) and $\log(g)=5.5$
(lower line)(${T_{\rm eff}=2700K}$, 
$\left[\case{{\rm M}}{{\rm H}}\right]=0.0$)}
\end{figure*}

\begin{figure*} 
\hbox{
}
\caption{\label{zvdw}
The  calculated \ion{Na}{1}~$\lambda\lambda8183,8195$ and \ion{K}{1}~$\lambda\lambda7665,7699$ 
doublets 
in vacuum for
$\left[\case{{\rm M}}{{\rm H}}\right]=0.5$ (upper line) and 
$\left[\case{{\rm M}}{{\rm H}}\right]=0.0$(lower line)
(${T_{\rm eff}=2700K}$, $\log(g)=5.5$)}
\end{figure*}

To demonstrate the various effects influencing the line profiles we will use the
\ion{Na}{1}~$\lambda\lambda8183,8195$ subordinate doublet and the
\ion{K}{1}~$\lambda\lambda7665,7699$ resonance doublet, since they are available
in the Keck-spectra (see Sec.  \ref{obsfit}).  Note that the synthetic line
profiles in this section are shown at their {\em vacuum} wavelengths.

Eq. (\ref{gammavdw}) shows that the vdW damping constant is a linear
function of the perturber density.  This implies that the major contribution to
the total vdW broadening will be due to the effect by the most abundant
perturbers, namely \ion{H}{1}, \ion{He}{1} and ${\rm H_2}$, because the
polarizabilities change much less than the relative concentrations of the
perturbing species (cf.~Tab.~\ref{perttab}).  In Fig.  \ref{diffvdw} we show the
influence of ${\rm H_2}$ on the profile of the subordinate
\ion{Na}{1}~$\lambda\lambda8183,8195$ doublet.  All the other perturbers listed
in Tab.~\ref{perttab} do not contribute significantly to the line width, the
profile is identical with the one which is only H--He--${\rm H_2}$ broadened.
This is valid even for metal abundances higher than solar.

In addition, Fig.~\ref{diffvdw} shows the effect of the correction factor
introduced in Eq.~(\ref{c6corr}) by artificially applying it to the alkali metal
sodium.  In this case the correction factor was only applied to demonstrate the
uncertainty which still is present in the vdW damping constant.  A final
decision on the validity of such a factor has to be made by comparing the
synthetic profiles with observations.

Changes of the absolute number density of each perturber will change the
damping constant.  We consider two ways to change the perturber densities:
changing the gravity or changing the metallicity.  Higher surface gravities will increase
the total pressure in every layer of the atmosphere.  This will also increase
the number density of every perturber for constant metallicity.  An example of
the gravity effect on the line widths is shown in Fig.~\ref{loggvdw} which
illustrates the influence of $\log(g)$ on the profile of the subordinate
\ion{Na}{1} doublet as well on the \ion{K}{1}~$\lambda\lambda7665,7699$
resonance doublet for model atmospheres with $\Teff=2700\,$K, solar metallicity
and a difference in the gravities of one order of magnitude.

On the other hand, an increase in metallicity at constant gravity will decrease
both the total pressure and the number density of the main perturbers,
\ion{H}{1}, \ion{He}{1} and ${\rm H_2}$.  It furthermore will bind hydrogen in
hydrides and water vapour.  Therefore, all these effects will make the line wings
narrower.  We demonstrate this in Fig.~\ref{zvdw}, with the same two doublets as
above.  The only difference in the models shown is half an order of magnitude in
the metallicity.

This introduces the problem that small changes in metallicity can be compensated
for by changes in the gravity, thus making the determination of both parameters
from the wings of strong lines alone questionable.  This means that abundance
analyses of cool dwarfs in general must be performed with caution and simple
techniques may lead to errors.  This will be investigated in much more detail in
subsequent work.

\subsection{Comparison to other methods}
\label{othmeth}

\begin{figure} 
\caption{\label{frakur}The calculated \ion{Na}{1}~$\lambda\lambda8183,8195$
doublet in vacuum with different 
broadening methods ($\log(g)=5.0$,  ${T_{\rm eff}=2700K}$, 
$\left[\case{{\rm M}}{{\rm H}}\right]=0.0$).
From top to bottom~: 
our method without the correction factor,
using Kurucz's damping constant per Hydrogen atom at 10\,000~K,
our method and  the correction factor of $10^{1.8}$ artificially applied,
the old method. See Sec. 3.2 
for details.}
\end{figure}

To demonstrate the improvement brought by the present
handling, we will briefly review other commonly used
methods: (1) the Uns{\"o}ld formula without
perturber correction used by \cite{allard90} and AH95, and
(2) using the precalculated damping constants per
hydrogen atom at 10\,000~K provided on CD-ROM~No.1~by 
\cite{cdrom1}.

The method (1) used an average broadening mechanism by taking the
total particle density instead of the density of every perturber and by not
distinguishing between the different perturbers, i.  e., not taking into account
the different polarizabilities and reduced masses, latter influencing
the relative velocity in Eq. (\ref{gammavdw}).  In addition, it used the
correction factor for all lines.

In Fig.  \ref{frakur} we compared now the same
\ion{Na}{1}~$\lambda\lambda8183,8195$ profiles as above calculated with the old
method with one calculated with the new method but the correction factor
artificially applied for comparison.  As can be seen there are clear differences
between the two methods.  The new method results in narrower lines since it
accounts for the different polarizabilities and reduced masses which decrease
$C_6$ in Eq.  (\ref{classc6}) or $\gamma_{\rm vdW}$ in Eq. (\ref{gammavdw}) 
compared to a $C_6$ and $\gamma_{\rm vdW}$ calculated as described above
and, therefore, also decreases the damping constants of Eq. 
(\ref{gammatot}).

In Fig \ref{frakur} we also plot the profiles using the damping constants
provided by Kurucz.  We find only small differences between Kurucz's vdW damping
constant and ours, if we omit the correction factor.  In test calculations we
also compared the use of natural damping constants provided by Kurucz with those
calculated with the classical approximation.  Since the total line width in
M~dwarfs is dominated by vdW broadening, it makes little observable
difference which treatment of natural damping is employed to model their
spectra.  We adopt our own methods in the following discussion.

We furthermore compared directly the damping constants obtained with our  
approximation with the corresponding values from Kurucz's list and found only
small differences for the vast majority of the lines above 6000~{\AA}.  The
largest differences arise from the correction factor we included for non-alkali
species.  For the UV region the differences are larger.  Here the hydrogenic
approximation for the absorber we used differs substantially from the method
Kurucz used.  This does not affect the spectra we present here.
In general, the photospheric UV flux from M~dwarfs is negligible.
However, a more thorough investigation must be performed for stars that 
are significantly hotter than M~stars.

\subsection{Uncertainties in the broadening mechanism of molecular lines}

\begin{figure*} 
\hbox{
}
\caption{\label{molcorr} Calculated molecular lines for 
$\log(g)=5.0$, $\left [\frac{{\rm M}}{{\rm H}} \right ]=+0.5$, ${ 
T_{\rm eff}=2600K}$. 
Dotted~: Without the artificial correction factor of $10^{1.8}$,
Solid~: With the correction factor applied.
Left panel~: Not rotationally broadened. Right panel~: rotationally broadened 
with
$10\kms$.
}
\end{figure*}

As can be seen from Figs.  \ref{molcorr} and \ref{mollin} the molecular
lines are much narrower than the atomic lines, both in the observations
(cf.  Sec.  \ref{molobs}) as well as in our models.  This is 
due to the fact that each molecular line is very weak compared to the
strong atomic lines. Therefore only the Gauss cores remain visible
in the spectra and the line width is only the line width
of the Gauss core. Nevertheless, the Voigt profiles need to
be included since they lift the Gauss cores significantly
without changing the line width
as corresponding comparisons showed.
In addition, the
molecular lines lie so densely together that only their cores can remain visible;
any wing contribution will effectively be blocked by the core of neighboring molecular
lines.  This means that the total width of molecular {\em features} is
not as much dominated by the wings of the vdW broadening as for the
atomic lines. 

In order to estimate the error resulting from our approximations, we performed
test calculations in which we omitted the correction factor from Eq.
(\ref{c6corr}) for molecules which results in a change in the interaction constant by
nearly 2 orders of magnitude.  In contrast to the atomic lines (cf. Sec. 
\ref{synprf} and Fig. \ref{diffvdw}), we found only small differences in the overall
appearance of molecular lines in the optical (see Fig.  \ref{molcorr}), since
only the line wings change and they are completely blended in the spectrum.

The most important result, however, was the influence of rotational broadening
on the appearance of the molecular lines which can be seen in 
Fig.  \ref{molcorr} as well.  As we
applied rotational broadening on our test spectra we found that even a small
rotational velocity will broaden the lines so much that the uncertainties in the
vdW broadening become unimportant.  Only the cores of the stronger molecular
lines remain visible.  Also the influence of the gravity and metallicity become
less important.
 We furthermore found that the effect of rotational broadening rises
 continuously with increasing rotational velocity.
 As a consequence, the molecular line `haze' reveals a possibility
 to accurately measure the rotational velocity as it is very sensitive to it
(see Sec. \ref{molobs}).

\section{Comparison of VB10 high-resolution Keck spectra with synthetic 
spectra} 
\label{obsfit} 

Our observation of VB10 was obtained with the HIRES echelle on the Keck
telescope during a run on 1995 March 12 under clear conditions. 
Our exposure time was 30 minutes with  0.8 arcsec seeing. 
The instrumental setup and data
reduction were very similar to those described by \cite{basri95}. The
wavelength setting was slightly modified to include the subordinate 
\ion{Na}{1}~$\lambda\lambda8183,8195$ doublet, the 
\ion{K}{1}~$\lambda\lambda7665,7699$ resonance doublet,
the \ion{Ca}{1}~$\lambda6573$ resonance line, 
the \ion{Ba}{1}~$\lambda7911$ resonance line
and 
the \ion{Fe}{1}~$\lambda7913$, \ion{Fe}{1}~$\lambda8662$ and \ion{Fe}{1}~$\lambda8689$ 
subordinate lines.
Inclusion of \ion{Na}{1}~
meant that we could not also observe the \ion{Rb}{1}~ line discussed by
\cite{basri95}. 

We consider models with any combination of $\log(g) = 4.0, 4.5, 5.0, 5.5$
and scaled solar metallicities 
$\left [\frac{{\rm M}}{{\rm H}} \right ]=-0.5, 0.0, +0.5$.
For the purpose of this paper, we adopt a fixed effective temperature of 
2700~K for VB10, derived from fitting  low-resolution spectra obtained 
by \cite{kirk93} and \cite{jones94} to low-resolution models \cite[see][]{schweitzer95}. 
It is not the purpose of this paper to investigate the influence of the
effective temperature on the line profile. This will be done
in subsequent work where the parameters will be verified or revised, if
necessary.
A rotational velocity of $8\kms$ was adopted, using
the method of \cite{basri95}. We reconsider this in section \ref{molobs}.
Due to the complexity of the spectra we performed the fits
and decided on their quality by eye.

\subsection{The sodium doublet}

\begin{figure*} 
\hbox{
}
\caption{\label{Na}The fit to the \ion{Na}{1}~$\lambda\lambda8183,8195$ doublet.
Solid~: Observed.
Dotted~: ${T_{\rm eff}=2700K}$, $\log(g)=5.0$, $\left [\frac{{\rm M}}{{\rm H}} 
\right ]=0.0$
(left) and
 ${T_{\rm eff}=2700K}$, $\log(g)=5.5$, $\left [\frac{{\rm M}}{{\rm H}} \right 
]=+0.5$ (right).
Both models with $8\kms$ rotationally broadened.
}
\end{figure*}

The \ion{Na}{1}~$\lambda\lambda8183,8195$ doublet lines are subordinate
transitions and, therefore, form mainly in the photosphere and should not be
significantly affected by the chromosphere of VB10.  Our best fits are the two
models with $\log(g)=5.0$ and $\left[\case{{\rm M}}{{\rm H}}\right]=0.0$ and
with $\log(g)=5.5$ and $\left[\case{{\rm M}}{{\rm H}}\right]=+0.5$ shown in Fig.
\ref{Na}.  There are no significant differences in the appearance of the two
model spectra, although the parameters differ substantially.  This demonstrates
the effect mentioned above, namely that an increase in metallicity decreases the
number density of the major perturbers, \ion{H}{1}, \ion{He}{1} and ${\rm H_2}$,
and thus partly cancels the effect of increasing the gravity, in particular for
strong lines.  This shows that one has to be very careful when analyzing M dwarf
spectra in detail.  
In our case here, we found good agreement between observation and
models for parameters within the intervals $5.0 < \log(g) < 5.5$
and $0.0 < \left[\case{{\rm M}}{{\rm H}}\right] < +0.5$.
We therefore use this as an error limit on our analyses.
For any parameter combination beyond these intervals we find
no reasonable agreement.

The synthetic line profiles show a deeper core than the observed lines. 
We calculated several models with a different microturbulence in order to 
``fill in'' the core but we found no significant improvement to the fit.
The remaining difference is most likely due to NLTE effects which might make 
line cores less deep. This will be investigated in detail in future work 
(see e.g., \cite{IAU176}).
NLTE effects will have to be considered as they are highly non-linear and are 
very likely to affect the cores of strong lines.

We also like to point out the numerous weak lines
in the wings of the atomic lines. These are mostly TiO lines which are
also visible in the observed spectrum and are not to be
considered as noise (see also below, Sec. \ref{molobs}).

\subsection{The potassium doublet}

\begin{figure*} 
\hbox{
}
\caption{\label{K} The fit to the \ion{K}{1}~$\lambda\lambda7665,7699$ doublet.
Solid~: Observed.
Dotted~: ${T_{\rm eff}=2700K}$, $\log(g)=5.0$, $\left [\frac{{\rm M}}{{\rm H}} 
\right ]=0.0$
(left) and
 ${T_{\rm eff}=2700K}$, $\log(g)=5.5$, $\left [\frac{{\rm M}}{{\rm H}} \right 
]=+0.5$ (right).
Both models with $8\kms$ rotationally broadened.
}
\end{figure*}

The \ion{K}{1}~$\lambda\lambda7665,7699$ lines show a clear core 
reversal (see Fig. \ref{K}). This is due to the chromospheric activity 
of VB10. The doublet is a resonance transition and thus will be strongly 
influenced by chromospheric activity. Since we are not treating the 
chromospheric effects here, our models can only reproduce the line wings 
which are formed deeper in the photosphere. 

As in the case of the sodium lines, both models fit equally well and a 
distinction between the two parameter sets of $\log(g)=5.0$ and 
$\left[\case{{\rm M}}{{\rm H}}\right]=0.0$ and of $\log(g)=5.5$ and 
$\left[\case{{\rm M}}{{\rm H}}\right]=+0.5$ is not readily possible (see 
Fig.\ \ref{K}). 
The fit itself is not as good as in the case of the sodium doublet 
as the models show many more molecular lines. 
This spectral region is partially contaminated by the telluric 
O$_2$ A bands present between 7595{\AA} and 7680{\AA} which we 
did not remove from the observed data. They are affecting the
observations by decreasing the continuum flux to a not readily
known extend.
In addition, this part of the spectrum is dominated by the $\varepsilon$-band of TiO
for which the line data are of lower quality (AHS96).
Nevertheless, the lines are so broad, that we still can fit
them as shown in Fig. \ref{K}.

\subsection{The calcium line}

\begin{figure*} 
\hbox{
}
\caption{\label{Ca} The fit to the \ion{Ca}{1}~$\lambda6573$ line.
Solid~: Observed.
Dotted~: ${T_{\rm eff}=2700K}$, $\log(g)=5.0$, $\left [\frac{{\rm M}}{{\rm H}} 
\right ]=0.0$
(left) and
 ${T_{\rm eff}=2700K}$, $\log(g)=5.5$, $\left [\frac{{\rm M}}{{\rm H}} \right 
]=+0.5$ (right).
Both models with $8\kms$ rotationally broadened.
Note H$_{\alpha}$ in emission.
}
\end{figure*}

The \ion{Ca}{1}~$\lambda6573$ resonance line shows also chromospheric 
features as can be seen in Fig.\ \ref{Ca}. 
The line is weak and very much
filled up and the line is hardly 
detectable without a specific identification. Calcium, 
in contrast to 
the two elements discussed above, 
is not an alkali metal. 
This means 
that the vdW damping constant is calculated including the correction 
factor introduced in Eq.~(\ref{c6corr}). 
Although it is hard to compare 
this weak line with models, we find that the correction factor is 
necessary in order to reproduce the observed line width with the same 
model parameters as used for the sodium and potassium lines. Without 
$C_6^{\rm corr}$ the theoretical line profile of the \ion{Ca}{1} lines is not 
broad enough, however, we cannot say how accurate our value for 
$C_6^{\rm corr}$ is. In order to make a more accurate determination of 
this factor, we will have to investigate much stronger non-alkali lines 
with clearer and broader wings. This will be done in a subsequent 
analysis. 

This line cannot be used to determine directly the parameters of VB10.
But it can be used to confirm them. As in the case of Sodium and Potassium
above, the two models with 
$\log(g)=5.0$ and $\left[\case{{\rm M}}{{\rm H}}\right]=0.0$ and
 $\log(g)=5.5$ and $\left[\case{{\rm M}}{{\rm H}}\right]=+0.5$
 produce equally good results.

\subsection{The barium line and the iron line at 7913 {\AA}}
\label{bafit}

\begin{figure*} 
\hbox{
}
\caption{\label{Ba} The fit to the \ion{Ba}{1}~$\lambda7911$ and 
\ion{Fe}{1}~$\lambda7913$ lines.
Solid~: Observed.
Dotted~: ${T_{\rm eff}=2700K}$, $\log(g)=5.0$, $\left [\frac{{\rm M}}{{\rm H}} 
\right ]=0.0$
(left) and
 ${T_{\rm eff}=2700K}$, $\log(g)=5.5$, $\left [\frac{{\rm M}}{{\rm H}} \right 
]=+0.5$ (right).
Both models with $8\kms$ rotationally broadened.
}
\end{figure*}

The fit to the \ion{Ba}{1}~$\lambda7911$ resonance line and 
the \ion{Fe}{1}~$\lambda7913$ line (which has 0.86 eV excitation energy)
is shown in Fig. \ref{Ba}.
Both lines are very weak but still visible and we find very
good agreement between observations and our two models with
$\log(g)=5.0$ and $\left[\case{{\rm M}}{{\rm H}}\right]=0.0$ and
$\log(g)=5.5$ and $\left[\case{{\rm M}}{{\rm H}}\right]=+0.5$.
As in the case of calcium above, we only used these
lines to confirm our range of fitting parameters,
since the line widths are not dominated by the
wings of the profiles.
They are dominated by their Gauss cores but the Voigt profiles
need to be included in order to lift the cores.

Both metals, barium and iron, are non-alakli metals as calcium above and,
therefore, treated with $C_6^{\rm corr}$ from Eq.~(\ref{c6corr}).
Although the lines are narrow, we find it also here
necessary to include 
such a correction factor in order to fit all lines
consistently. However, we cannot make any accurate determinations
of $C_6^{\rm corr}$, which still has to be done with stronger lines.

We also want to point out the very good fit of the surrounding molecular
lines (see also Sec. (\ref{molobs})).

\subsection{The iron lines at 8662 and 8689 {\AA}}

\begin{figure*} 
\hbox{
}
\caption{\label{Fe} The \ion{Fe}{1}~$\lambda8662$ and 
\ion{Fe}{1}~$\lambda8689$ lines.
Solid~: Observed.
Dotted~: ${T_{\rm eff}=2700K}$, $\log(g)=5.0$, $\left [\frac{{\rm M}}{{\rm H}} 
\right ]=0.0$
(left) and
 ${T_{\rm eff}=2700K}$, $\log(g)=5.5$, $\left [\frac{{\rm M}}{{\rm H}} \right 
]=+0.5$ (right).
Both models with $8\kms$ rotationally broadened.
There has been an offset applied to the observed spectrum. See text for details.
}
\end{figure*}

The two subordinate \ion{Fe}{1}~$\lambda8662$ and \ion{Fe}{1}~$\lambda8689$ 
lines visible in the observations are shown in Fig. \ref{Fe}.
The lines are not very strong, yet they can be clearly identified.
This spectral region is strongly contaminated by
terrestial absorption features which reduce the total flux and suppress 
the molecular pseudo continuum.
Therefore, we did not try to fit the lines in detail.
Instead we show them with an offset to demonstrate the
resemblance of the line shapes of the two \ion{Fe}{1} lines. 

We cannot use these lines to determine the parameters of VB10.
As in the cases above we used models with 
$\log(g)=5.0$ and $\left[\case{{\rm M}}{{\rm H}}\right]=0.0$ and
$\log(g)=5.5$ and $\left[\case{{\rm M}}{{\rm H}}\right]=+0.5$
and found consistency within the given constraints.
Again, calculating the vdW damping constant
including $C_6^{\rm corr}$ (see also Sec. (\ref{bafit}), 
results in consistent synthetic
spectra. But due to the uncertainties
in the observed spectrum we did not try to determine the accuracy
of $C_6^{\rm corr}$ here either.

We want to point out the `edge' at approximately 8671 {\AA} in our models,
which is due to an abrupt end of a VO band.
In our models we can treat VO only with the JOLA method (AH95, AHS96)
since no line list is yet available. 
Since the JOLA method will produce these edges, it is not
suitable for computing high-resolution spectra.
We point out that the \ion{Fe}{1}~$\lambda8662$ line
is slightly contaminated by a \ion{Ca}{2}~$\lambda8662$ line.

\subsection{Molecular lines}
\label{molobs}

\begin{figure*} 
\hbox{
}
\caption{\label{mollin} Molecular lines.
Solid~: Observed.
Dotted~: ${T_{\rm eff}=2700K}$, $\log(g)=5.0$, $\left [\frac{{\rm M}}{{\rm H}} 
\right ]=0.0$
(left) and
 ${T_{\rm eff}=2700K}$, $\log(g)=5.5$, $\left [\frac{{\rm M}}{{\rm H}} \right 
]=+0.5$ (right).
Both models with $8\kms$ rotationally broadened.
}
\end{figure*}

As mentioned before, the high-resolution spectra are full with a large 
number of weak molecular lines. For TiO, we use the semi-empirical line 
list of \cite{TiOjorg}, which will not 
reproduce most of the TiO lines at their exact wavelength or their proper gf-value. 
But nevertheless many of the lines agree very well with  observations.
We found an excellent agreement
with models of the two parameter sets of 
$\log(g)=5.0$ and $\left[\case{{\rm M}}{{\rm H}}\right]=0.0$ and
$\log(g)=5.5$ and $\left[\case{{\rm M}}{{\rm H}}\right]=+0.5$
as can be seen in Fig. \ref{mollin}.
Because of the strong influence of rotational broadening, we only use
the molecular lines to confirm the parameters and not to 
derive them directly.
On the other hand, we can derive the rotational velocity a 
second time by using its strong influence on the profile
of the molecular lines.
We found that rotational velocities 
between 7 and $9\kms$ produce very similar fits,
whereas applying values beyond this
interval result in worse fits.
Therefore, we can confirm a rotational velocity of $8\pm1 \kms$.

\section{Summary and Conclusions}
\label{results}

We found that the line widths caused by pressure broadening due to van der Waals
damping are very sensitive to both the gravity and the metallicity.  Therefore
fitting the lines of high resolution spectra provides a very important tool to
determine these two parameters.  Nevertheless, since both parameters influence the
line widths in the similar ways, one has to be very careful with the
analysis.

From the fits to the lines described above we estimate that VB10 has a 
gravity in the range $5.0 < \log(g) < 5.5$ and a metallicity $ 0.0 < 
\left[\case{{\rm M}}{{\rm H}}\right] < 0.5$ based on line broadening 
considerations alone with an adopted effective temperature of
${T_{\rm eff}}=2700{\rm K}$ \cite[]{schweitzer95}. 
We will investigate the influence of the effective temperature on the
line profiles in subsequent work.
However, $\log(g)=5.5$ as well as $\left[\case{{\rm M}}{{\rm H}}\right] = +0.5$
are rather high values and not expected for VB10 
\cite[cf. e.g.][]{henry93}, so that the lower part
of the interval is physically more likely.

Our calculations result in damping constants for alkali 
metals that agree satisfactorily well with the Keck observations.  
The damping constants for non-alkali metals include the correction 
factor of $C_6^{\rm corr}=10^{1.8}$ and lead to line 
profiles which are consistent with the data.
In order to obtain a better value for the correction
factor, observations of strong non-alkali lines are required.
We calculated synthetic spectra outside the observed
spectral range available in this paper in order to find such
strong lines, but the only line is
a moderately strong \ion{Fe}{1}$\lambda{8829}$ line besides the 
\ion{Rb}{1}$\lambda7800$ and \ion{Rb}{1}$\lambda7948$ lines.
The iron lines are stronger in hotter M~dwarfs which would be more
suitable for testing the vdW broadening of iron lines.
We also find
that our damping constants for atomic lines are in broad agreement with the values provided by
Kurucz.
No error should be expected from uncertainites in the
oscillator stengths of the atomic lines since they are
much smaller than any other uncertainties and are well known for the light
element lines that we have mostly considered here.

Our analysis of the Keck spectra yields values for the gravity which are consistent with the
results of interior calculations for VB10.  The ${T_{\rm eff}}$,
$\left[\case{{\rm M}}{{\rm H}}\right]$ and $\log(g)$ combinations agree reasonably
well with evolutionary models for M dwarfs \cite[]{burrow93,baraffe95}.
Larger damping constants, e.g.  as a result from a different treatment of line
broadening, would immediately imply a too low $\log(g)$. 

There are still discrepancies between models and observations in
some molecular line dominated regions of the spectrum.  As mentioned we suspect this
to be due to the molecular line data.  This will be improved in subsequent work.
We also will investigate the influence of Non-LTE effects on M~dwarf spectra as
well as chromospheric features.
In addition, 
observations of different strong
atomic lines in spectral regions where molecular absorption is not
dominating and blending is less severe \nocite{jones96}
(e.g. beyond 1.18$\mu$m, see also Jones et al., 1996),
will clarify the calculations of the damping constants and
the use of correction factors.

\section*{Acknowledgements}

We would like to express our thanks to J. Krautter, S. Starrfield and
I. Appenzeller who made this collaboration possible and supported
it.

We also thank the referee, H.R.A Jones, for his very helpful
comments.

This research has 
been partially supported by NASA LTSA grants NAGW 4510 and
NAGW 2628 and NASA ATP grant NAG 5-3067 to ASU as well as by
NSF grant AST-9217946 to WSU.  Parts of the model calculations have 
been performed on the Cray C90 of the San Diego Supercomputer Center 
and on the IBM SP2 of the Cornell Theory Center, supported by the NSF.
We thank them for a generous allocation of computer time.

A. Schweitzer was partially supported by a BAF{\"o}G grant.

G. Basri wishes to thank the University of California for partial travel support to the
Keck Observatory.

\end{document}